\documentclass[]{aa}
\usepackage{txfonts}
\usepackage{graphicx}

\begin{document}

\title{The formation of spiral arms and rings in barred galaxies}
\author{M. Romero-G\'omez\inst{1,2}, E. Athanassoula\inst{1},
J.J. Masdemont\inst{3}, \and C. Garc\'{\i}a-G\'omez\inst{2}}
\institute{LAM, Observatoire Astronomique de Marseille Provence, CNRS, 
2 Place Le Verrier, 13248 Marseille C\'edex 04, France
\and D.E.I.M., Universitat Rovira i Virgili, Campus Sescelades, Avd. dels Pa\"{\i}sos
Catalans 26, 43007 Tarragona, Spain
\and I.E.E.C \& Dep. Mat. Aplicada I, Universitat Polit\`ecnica de
Catalunya, Diagonal 647, 08028 Barcelona, Spain}
\offprints{M. Romero-G\'omez, \email{merce.romero@urv.cat}}
\date{Received 19 March}
\abstract{In this and in a previous paper (Romero-G\'omez et al. \cite{rom06})
we propose a theory to explain the formation of both spirals and
rings in barred galaxies using a common dynamical framework. It is based on the
orbital motion driven by the unstable equilibrium points of
the rotating bar potential. Thus, spirals, rings and pseudo-rings are
related to the invariant manifolds associated to the periodic orbits
around these equilibrium points. We examine the parameter space of three barred galaxy 
models and discuss the formation of the different 
morphological structures according to the properties of the bar model. We also study 
the influence of the shape of the rotation curve in the outer 
parts, by making families of models with rising, flat, or falling rotation curves 
in the outer parts. The differences between spiral and ringed structures arise from 
differences in the dynamical parameters of the host galaxies. The results presented 
here will be discussed and compared with observations in a forthcoming paper.

}
\authorrunning{M. Romero-G\'omez et al.}
\titlerunning{The formation of rings and spirals in barred galaxies}
\maketitle

\section{Introduction}

Bars are a very common feature of disc galaxies. In a sample of $186$
spirals drawn from the Ohio State University Bright Spiral Galaxy
Survey, Eskridge et al. (\cite{esk00}) find that in the near infrared $56\%$ of the
galaxies are strongly barred, while an additional $6\%$ are weakly 
barred. Only $27\%$ can be classified as non-barred. A large 
fraction of barred galaxies show two clearly defined spiral
arms (e.g. Elmegreen \& Elmegreen \cite{elm82}), often departing from the end 
of the bar at nearly right angles. This is the case for instance in NGC~1300, 
NGC~1365 and NGC~7552. Deep exposures, moreover, show that these arms wind around 
the bar structure and extend to large distances from the centre (see for instance 
Sandage \& Bedke \cite{san94}). Almost all researchers agree that
spiral arms and rings are driven by the gravitational field of the
galaxy (see Toomre \cite{too77} and Athanassoula \cite{ath84}, for reviews). 
In particular, spirals are believed to be density waves in 
a disc galaxy (Lindblad \cite{lind63}). Toomre (\cite{too69}) found that the spiral 
waves propagate towards the principal Lindblad resonances of the galaxy, where they 
damp down, and thus concludes that long-lived spirals need some replenishment. 

There are essentially three different possibilities for a spiral wave to be replenished. 
First, it can be driven by a companion or satellite galaxy. A direct, relatively slow 
and close passage of another galaxy can form trailing shapes (e.g. Toomre \cite{too69}; 
Toomre \& Toomre \cite{too72}; Goldreich \& Tremaine \cite{gol78}, \cite{gol79}; 
Toomre \cite{too81} and references therein). They can also be excited by the 
presence of a bar. Several studies have shown that a rotating bar or oval can 
drive spirals (e.g. Lindblad \cite{lind60}; Toomre \cite{too69}; Sanders \& Huntley 
\cite{san76}; Schwarz \cite{sch79}, \cite{sch81}; Huntley \cite{hun80}). Athanassoula 
(\cite{ath80}) studied the self-consistent quasi-stationary response of a galaxy 
composed of both gas and stars to a growing and rotating bar. She showed 
that the response of both components is bar-like up to corotation, where it turns into 
a trailing two-armed spiral, ending approximately at the Outer Lindblad Resonance (OLR). 
The third alternative, proposed by Toomre (\cite{too81}), is the swing amplification
feedback cycle. This starts with a leading wave propagating from the
center towards corotation. In doing so, it unwinds and then winds in
the trailing sense, while being very strongly amplified. This trailing wave will propagate
towards the center, while a further trailing wave is emitted at CR and
propagates outwards, where it is dissipated at the OLR.
The inwards propagating trailing wave, when reaching the center
will reflect into a leading spiral which will propagate outwards
towards the CR, thus closing the feedback cycle. Note that, if there is an Inner
Lindblad Resonance (ILR), the wave propagating inwards is damped at that resonance and 
the cycle is cut.
 
Strongly barred galaxies can also show prominent and spectacular rings or partial rings. 
The origin of such morphologies has been studied by Schwarz (\cite{sch81}, \cite{sch84}, 
\cite{sch85}), who followed the response of a gaseous disc galaxy to a bar perturbation. He
proposed that ring-like patterns are associated to the principal orbital resonances, 
namely ILR, CR, and OLR. The ILR would be responsible for the nuclear rings, CR would
be associated with the inner rings, which are indicated by an {\sl r} in the de Vaucouleurs 
classification, and the OLR would be the origin of the outer rings, which are indicated by 
an {\sl R} preceding the Hubble type. Nuclear rings are small rings of star formation
often found near the centres or nuclei of early-type barred galaxies, but we will not discuss 
these structures here. Inner rings are the well-defined rings that encircle the bars 
of barred galaxies. They are sometimes active sites of star formation. Outer 
rings are the larger, more diffuse rings, about twice the size of the bar. When these 
structures are incomplete, we designate them with the term 
pseudo-rings. There are different types of outer rings. Buta (\cite{but95})
classified them according to the relative orientation of the ring and bar major axes. 
If these two axes are perpendicular, the shape of the ring is similar to an ``8'' 
and the outer ring is classified as $R_1$. If the two axes are parallel, the outer 
ring is classified as $R_2$. Finally, if both types of rings are present in the galaxy,
the outer ring is classified as $R_1R_2$.

In this paper, we develop further the basic idea that we proposed in a
previous paper (Romero-G\'omez et al. \cite{rom06}, hereafter Paper
I). Rings and spiral arms are the result of the orbital motion 
driven by the invariant manifolds associated to periodic orbits around
unstable equilibrium points. In Paper I we gave a complete and
thorough description of the dynamical framework and, as an example,
specifically calculated the orbital evolution of a particular type 
of $rR_1$ ringed galaxy. In this paper, we construct families of
models based on simple, yet realistic, barred galaxy potentials. In
each family, we vary one of the free parameters of the potential
and keep the remaining fixed. For each model, we compute numerically the 
orbital structure associated to the invariant manifolds.
In this way, we are able to study the influence of each model parameter on the
global morphologies delineated by the invariant manifolds.

In Sect. \ref{sec:mod}, we present the equations of motion and the galactic models
that will be used for the computations. In Sect. \ref{sec:inv}, we give a brief
description of the invariant manifolds and describe their role in the
transfer of matter within the galaxy. In Sect. \ref{sec:res}, we show the
different morphologies that result from the computations of each model and in 
Sect. \ref{sec:con} we briefly summarize. Applications to real galaxies, as well as
comparisons with observations and with other theoretical work (Schwarz \cite{sch81}, \cite{sch84},
\cite{sch85}; Kaufmann \& Contopoulos \cite{kau96}; Patsis \cite{pat06}; Voglis, Stavropoulos \& 
Kalapotharakos \cite{vog06}) will be given in 
paper III of this series.

\section{Models}
\label{sec:mod}

\subsection{Equations of motion and equilibrium points}
\label{sec:eq}
We model the potential of a barred galaxy as the superposition of three
components, two of them axisymmetric and the third bar-like. This
latter component rotates anti-clockwise with angular velocity ${\bf
\Omega_p}=\Omega_p{\bf z}$, where $\Omega_p$ is a constant pattern
speed\, \footnote{Bold letters denote vector notation. The vector {\bf z} is a unit
vector.}. The equations of motion in 
this potential in a frame rotating with angular speed ${\bf \Omega_p}$ in vector form are
\begin{equation}\label{eq-motvec}
{\bf
\ddot{r}=-\nabla \Phi} -2{\bf (\Omega_p \times \dot{r})-  \Omega_p \times
(\Omega_p\times r)},
\end{equation}
where the terms $-2 {\bf \Omega_p\times \dot{r}}$ and $-{\bf \Omega_p \times
(\Omega_p\times r)}$ represent the Coriolis and the centrifugal
forces, respectively, and ${\bf r}$ is the position vector. 

Following Binney \& Tremaine (\cite{bint87}), we take
the dot product of Eq. (\ref{eq-motvec}) with
${\bf \dot{r}}$, and by rearranging the resulting equation, we obtain
$$
\frac{d E_J}{dt}=0,
$$
where 
$$
E_J \equiv \frac{1}{2} {\bf\mid\dot{r}\mid}^2 + \Phi -\frac{1}{2}\mid {\bf\Omega_p
\times r}\mid^2.
$$
The quantity $E_J$ is a constant of the motion and is known as the Jacobi
integral, or as the Jacobi constant. Note that it
is the sum of $\frac{1}{2} {\bf\dot{r}^2} + \Phi$, which is
the energy in a non-rotating frame, and of the quantity 
$$ -\frac{1}{2}\mid {\bf\Omega_p\times
r}\mid^2=-\frac{1}{2}\Omega_p^2\,(x^2+y^2),$$ 
which can be
thought of as the ``potential energy'' to which the centrifugal
``force'' gives rise. Thus, if we define an effective potential
$$\Phi_{\hbox{\scriptsize eff}}=\Phi-\frac{1}{2}\Omega_p^2\,(x^2+y^2),
$$ Eq. (\ref{eq-motvec}) becomes
$$
{\bf \ddot{r}=-\nabla \Phi_{\hbox{\scriptsize eff}}} -2{\bf (\Omega_p \times \dot{r})},
$$
and the Jacobi constant is 
$$
E_J = \frac{1}{2} {\bf\mid \dot{r}\mid} ^2 + \Phi_{\hbox{\scriptsize eff}},
$$
which, being constant in time, can be considered as the energy in the
rotating frame.

Setting $x_1=x,\quad x_2=y,\quad x_3=\dot{x}$, and $\quad
x_4=\dot{y}$, we can write the components of the equations of motion as 
\begin{equation}\label{eq-first}
\left \{ \begin{array}{rclcl}
\dot{x_1} & = & f_1(x_1,x_2,x_3,x_4) & = & x_3\\
\dot{x_2} & = & f_2(x_1,x_2,x_3,x_4) & = & x_4\\
\dot{x_3} & = & f_3(x_1,x_2,x_3,x_4) & = & 2\Omega_p\, x_4-\Phi_{x}\\
\dot{x_4} & = & f_4(x_1,x_2,x_3,x_4) & = & -2\Omega_p\, x_3-\Phi_{y},\\
\end{array} \right.
\end{equation}
where we have defined $\Phi_x=\frac{\partial \Phi_{\hbox{\scriptsize
eff}}}{\partial x}$ and
$\Phi_y=\frac{\partial \Phi_{\hbox{\scriptsize eff}}}{\partial
y}$. Note that we restrict ourselves to the $z=0$ plane because the
motion in the vertical direction essentially consists of an uncoupled
harmonic oscillator and does not affect the motion in this plane (see
Paper I).

The surface $\Phi_{\hbox{\scriptsize eff}}=E_J$ is called the zero
velocity surface, and its intersection with the $z=0$ plane gives the
zero velocity curve. All regions in which $\Phi_{\hbox{\scriptsize
eff}}>E_J$ are forbidden to a star with this energy, and are thus called 
forbidden regions. The zero velocity curve also defines two different regions,
namely, an exterior region and an interior one that contains the bar. The
interior and exterior regions are connected via the equilibrium points (see 
Fig. 2b of Paper I). 

The bar component is rotating as a solid body with a constant pattern 
speed $\Omega_p$. For our calculations we place ourselves in a frame of 
reference corotating with the bar, and the bar semi-major axis is located
along the $x$ axis. In this rotating frame we have five equilibrium
Lagrangian points. Three of these points are stable, namely $L_3$,
which is placed at the centre of the system, and $L_4$ and $L_5$,
which are located symmetrically on the $y$ axis. The two equilibrium
points left, $L_1$ and $L_2$, are unstable and are located
symmetrically on the $x$ axis. The position of the corotation radius
will be determined by the free parameter $r_L$, that is, the distance
from the centre to $L_1$, or, equivalently, to $L_2$, because the
model is symmetric with respect to both $x$ and $y$ axes. In our case,
we choose $r_L$ to be the distance from the centre to $L_1$. The
pattern speed is related to $r_L$ through the expression
\begin{equation}\label{eq:omegap}
\Omega_p^2 = r_L\left(\frac{\partial \Phi(r)}{\partial r}\right)_{r_L},
\end{equation}
where $\Phi(r)$ is the potential on the equatorial plane. 

\subsection{Galaxy models and free parameters}
\label{sec:model}
In this section we describe the potentials that we use
to model a barred galaxy. Our basic model will be the
one introduced by Athanassoula (\cite{ath92a}) in a thorough study of the orbital
structure of bars. The axisymmetric component consists of the superposition of a disc 
and a spheroid, whose basic parameters are determined so that the rotation
curve of the galactic model has the desired characteristics. The disc is modelled as 
a Kuzmin-Toomre disc (Kuzmin \cite{kuz56}; Toomre \cite{too63}) of surface density
\begin{equation}\label{eq:kuz}
\sigma(r) = \frac{V_d^2}{2\pi r_d}\left(1+\frac{r^2}{r_d^2}\right)^{-3/2}.
\end{equation}
The parameters $V_d$ and $r_d$ set the scales of the velocities and radii, respectively. 
The spheroid is modelled using a density distribution of the form
\begin{equation}\label{eq:sph}
\rho(r)=\rho_b\left(1+\frac{r^2}{r_b^2}\right)^{-3/2},
\end{equation}
where $\rho_b$ and $r_b$ determine its
concentration and scale length. Spheroids with high concentration have 
high values of $\rho_b$ and small values of $r_b$, the opposite being
true for spheroids of low concentration. Although we give two separate axisymmetric 
components, it is important to note that, in fact, what matters in this study is only 
the total axisymmetric rotation curve and not its decomposition into components. 

The bar component is described by a Ferrers (\cite{fer77}) ellipsoid whose density 
distribution is described by the expression:
\begin{equation}
\left\{\begin{array}{lr}
\rho_0(1-m^2)^n & m\le 1\\
 0 & m\ge 1,
\end{array}\right.
\label{eq:Ferden}
\end{equation}
where $m^2=x^2/a^2+y^2/b^2$. The values of $a$ and $b$
determine the shape of the bar, $a$ being the length of the semi-major
axis, which is placed along the $x$ coordinate axis, and $b$ being the
length of the semi-minor axis. The parameter $n$ measures the degree of concentration 
of the bar. High values of $n$ correspond to a high concentration, while a 
value of $n=0$ is the extreme case of a constant density bar. The parameter 
$\rho_0$ represents the bar central density. For these models, the 
quadrupole moment of the bar is given by the expression
$$Q_m=M_b(a^2-b^2)/(5+2n), $$
where $M_b$ is the mass of the bar, equal to
$$ M_b=2^{(2n+3)}\pi ab^2 \rho_0 \Gamma(n+1)\Gamma(n+2)/\Gamma(2n+4) $$
and $\Gamma$ is the gamma function. 
Throughout this paper we will use the following system of units: For the mass unit 
we take a value of $10^6 M_{\odot}$, for the length unit a value of $1$ kpc and 
for the velocity unit a value of $1\,\,\rm{km}\,\rm{s}^{-1}$. Using these values, the 
unit of the Jacobi constant will be $1\,\,\rm{km}^2\,\rm{s}^{-2}$.

For reasons of continuity and to allow in Paper III a comparison of our results with 
the results of gas flow in these models (Athanassoula \cite{ath92b}), we will use the same 
numerical values for the model parameters as in Athanassoula (\cite{ath92a}). We will 
hereafter refer to this model as model A. It has essentially four free parameters which 
determine the dynamics in the bar region. The axial ratio $a/b$ and the quadrupole moment 
(or mass) of the bar $Q_m$ (or $M_b$), will determine the strength of the bar. The third 
parameter is the bar angular velocity, or pattern speed, determined by the Lagrangian radius 
$r_L$ (Eq. \ref{eq:omegap}). The last free parameter is the central concentration of the model 
$\rho_c=\rho_b + \rho_0$. As already mentioned, the basic values for the free parameters 
are set as in run 001 of Athanassoula (\cite{ath92a}): $a/b=2.5$, $r_L=6$, $Q_m=4.5\times 
10^4$, $\rho_c=2.4\times 10^4$. Then a range around each of these values is explored. The 
axisymmetric component is fixed by setting a maximum disc circular velocity of $164.204$ at 
$r=20$, and $r_b$ is determined by fixing the total mass of the spheroid and bar components 
within $r=10$ to $4.87333\times 10^4$, while fixing the combined central density of the bar and 
bulge to $\rho_c$. The length of the bar is fixed to 5. Results on the orbital structure 
underlying spirals and rings in such models will be presented in Sect. \ref{sec:fer01}.

So far, we have followed exactly the model of Athanassoula (\cite{ath92a}). Nevertheless,
in this paper we are interested in spirals and rings which occur 
beyond or around CR, contrary to Athanassoula (\cite{ath92a}, \cite{ath92b}), who
concentrated on orbital structure and gas flow within and around the bar region. We will, 
therefore, consider a further option, namely whether the rotation curve in the region 
beyond CR is flat (model F), or somewhat rising (model R), or somewhat falling (model D). 
This will be achieved by considering the same axisymmetric components as in model A, but 
with different values of the parameters. The different shapes of the rotation curve in the 
outer parts are obtained by giving different maximum disc circular velocities at a fixed 
radius. Thus, model R, with a rising rotation curve, has a maximum disc circular velocity of 
$164.204$ at $r=20$. Model F, with a flat rotation curve in the outer parts, has a 
maximum disc circular velocity of $100$ at $r=20$, while model D, with a decreasing 
rotation curve in the outer parts, has a maximum disc circular velocity of $10$ at $r=20$. 
The spheroid scale length is determined as in model A. In this case, we use an 
inhomogeneous Ferrers bar ($n=1$) with axial ratio $a/b=2.5$ and a semi-major axis 
$a=5$, and a central concentration $\rho_c=0.05 \times 10^4$. The basic values for the
remaining free parameters are taken to be $r_L=6$ and $Q_m=4.5 \times 10^4$. As for
the previous model, a range around each of these values is explored. Results using these 
models are presented in Sect. \ref{sec:ferRFD}.

The Ferrers bars are realistic models of bars and have been widely used so far in orbital 
structure studies within and in the immediate neighbourhood of bars (e.g. Athanassoula et 
al. \cite{ath83}; Pfenniger \cite{pfe84}, \cite{pfe87}, \cite{pfe90}; Skokos, Patsis \& 
Athanassoula \cite{sko02a}, \cite{sko02b}). They contain parameters with physical meaning, 
such as the bar mass or axial ratio, that can be obtained from, or compared to, observations. 
They have, however, one disadvantage, namely that in models using such bars the ratio of
the non-axisymmetric component of the force to the axisymmetric component of the force decreases 
very abruptly beyond a certain radius so 
that the axisymmetric component dominates in the outer regions. This is of no importance if 
one is interested in the orbital structure or the gas flow in the bar region, but in studies 
like this one, where one is interested in the region outside the bar, this may introduce a 
bias, since models with high nonaxisymmetric forces beyond CR will not be studied. In order 
to remedy this, we introduce two other models, also often used in the literature, which have 
an ad hoc bar potential, i.e. a potential that is not associated with a particular 
density distribution. Ad hoc models have some disadvantages. They are simple mathematical 
expressions for the potential and do not originate from a realistic density distribution. 
This means that the corresponding density distribution may have some undesired features, e.g. 
for very strong nonaxisymmetric perturbations the local density could even be negative. 
Furthermore, they do not contain simple parameters that can be directly and straightforwardly 
associated to observable quantities, like the bar length, mass, or axial ratio. Most of them 
are of the form $\epsilon A(r)\cos(2\theta)$, i.e. contain no $\cos(m\theta)$ terms with $m>2$. 
This means that the parameter $\epsilon$ is associated with the mass of the bar and that there 
is no parameter to regulate its axial ratio. Despite all these shortcomings, ad hoc potentials 
have been widely used because they have the important advantage of being adaptable to the 
problem at hand, i.e. with a proper choice of the function $A(r)$ one can obtain a 
potential with the desired properties, for example, in our case, a potential with an important 
$m=2$ contribution between CR and OLR.

The first ad hoc potential we use has the form
\begin{equation}
\Phi(r,\theta)=-\frac{1}{2}\epsilon v_0^2\cos(2\theta)\left\{{\begin{array}{ll}
\displaystyle 2-\left(\frac{r}{\alpha}\right)^n, & r\le \alpha\rule[-.5cm]{0cm}{1.cm}\\
\displaystyle \left(\frac{\alpha}{r}\right)^n, & r\ge \alpha.\rule[-.5cm]{0cm}{1.cm}
\end{array}}\right.
\label{eq:adhoc1}
\end{equation}
The parameter $\alpha$ is a characteristic length scale of the bar potential and 
$v_0$ is a characteristic circular velocity. The parameter $\epsilon$ is a free parameter 
related to the bar strength. Dehnen (\cite{deh00}) used this 
potential with $n=3$ to study the effect of the OLR of the bar of our
Galaxy on the local stellar velocity distribution. Fux (\cite{fux01})
uses the same bar potential to model the Galactic bar and to study the
order and chaos in the disc. Nevertheless, since the characteristics of the potential with 
$n=3$ are very similar to Ferrers potentials, i.e. the force decreases very
abruptly at large radii, we will use here $n=0.75$ to avoid this. We will couple
this bar with an axisymmetric part given by Eq. \ref{eq:kuz} and \ref{eq:sph} and
we will consider three different slopes of the rotation curve in the outer parts, like
we did for the previous models. In all three models, the length scale of the bar is set 
to $\alpha=5$ and the circular velocity $v_0$ is set to $200$. The basic value for the 
corotation radius is set to $r_L=6.0$, and for the bar strength to $\epsilon=0.15$ 
and a range around each of these values is explored. The spheroid mass is fixed at 
$M_{bul}=3\times 10^4$, and the scale length is calculated accordingly. In model R', which 
has a rising rotation curve, the disc parameters are fixed so that the maximum circular 
velocity is $164.204$ at $r=10$. In model F', which has a flat rotation curve in the outer 
parts, we fix the values of $r_d$ and $V_d$ so that the maximum circular velocity is 
$164.204$ at $r=50$. Model D', which has a decreasing rotation curve in the outer parts, 
has a maximum disc circular velocity of $164.204$ at $r=100$. Results using these models 
are presented in Sect. \ref{sec:dehRFD}.

We also used in our computations the bar potential given by the expression:
\begin{equation}\label{eq:adhoc2}
\Phi(r,\theta)=\hat{\epsilon}\sqrt{r}(r_1-r)\cos(2\theta),
\end{equation}
where $r_1$ is a characteristic scale length of the bar potential, which we will take 
for the present purposes to be equal to $20$. The parameter $\hat{\epsilon}$ is related 
to the bar strength. This type of model has already been widely used in studies of bar 
dynamics (e.g. Barbanis \& Woltjer \cite{bar67};  Contopoulos \& Papayannopoulos 
\cite{con80}; Contopoulos \cite{con81}; Athanassoula \cite{ath90}).  We will couple this 
bar with an axisymmetric part given by Eq. \ref{eq:kuz} 
and \ref{eq:sph}. We will consider two reference models noted with the names of 
models R'' and D'', again each with a different slope of the rotation curve in the outer 
parts. The strength parameter is set to $\hat{\epsilon}=100$ and the Lagrangian radius is 
$r_L=6$. The spheroid mass is $M_{bul}=3\times 10^4$, and the spheroid scale length is 
calculated accordingly. The disc parameters in model R'' are set so that the maximum disc 
circular velocity is $164.204$ at $r=30$, which corresponds to a rotation curve rising 
in the outer parts, while for model D'' they are obtained by fixing a 
maximum circular velocity of $164.204$ at $r=10$, which corresponds to a rotation 
curve falling in the outer parts. 

\section{Invariant manifolds and transfer of matter}
\label{sec:inv}
In this section, we give a simplified description of the dynamics
around the unstable equilibrium points. Readers interested in a more
thorough description, in particular of the Lyapunov periodic orbits and of
the corresponding invariant manifolds, are referred to Paper I and to 
references therein. Here, we also 
define the heteroclinic and homoclinic orbits and describe the role they play 
in the transfer of matter within the galaxy. We will use these ideas to compare
our results with observations in the forthcoming Paper III.

In Sect. \ref{sec:eq}, we mentioned that the equations of motion 
(Eq. \ref{eq-first}) have five equilibrium points, three of which are linearly
stable, namely $L_3$, $L_4$, and
$L_5$, and two unstable, namely $L_1$ and $L_2$. Around the equilibrium points 
there exist families of periodic orbits, e.g. around the central 
equilibrium point the well-known $x_1$ family of periodic orbits which is 
responsible for the bar structure. Around each unstable equilibrium point
 there also exists a family of periodic orbits, known as the family of 
Lyapunov orbits (Lyapunov \cite{lya49}), which, at low energy levels, are 
unstable and become stable 
only at high energies (Skokos et al. \cite{sko02a}). We are interested in this 
family only in the range of energies where the periodic orbits are unstable. 
In this range, the size of the periodic orbits remains small and they stay in 
the vicinity of the equilibrium point. For a given energy level within this 
range, two stable and two unstable sets of asymptotic orbits emanate from the 
periodic orbit and they are known as the stable and unstable invariant 
manifolds, respectively. We denote by $W^s_{\gamma_i}$ the stable invariant 
manifold associated to the periodic orbit $\gamma$ around the equilibrium 
point $L_i,\, i=1,2$. This stable invariant manifold is the set of 
orbits that tends to the periodic orbit asymptotically. In the same way we 
denote by 
$W^u_{\gamma_i}$ the unstable invariant manifold associated to the periodic 
orbit $\gamma$ around the equilibrium point $L_i,\, i=1,2$. This unstable 
invariant manifold is the set of orbits that departs asymptotically from the 
periodic orbit (i.e. orbits that tend to the Lyapunov orbits when the time 
tends to minus infinity). Since the invariant manifolds extend well beyond the 
neighbourhood of the equilibrium points, they can be responsible for 
global structures. 
\begin{figure*}
\centering
\includegraphics[scale=0.5,angle=-90.]{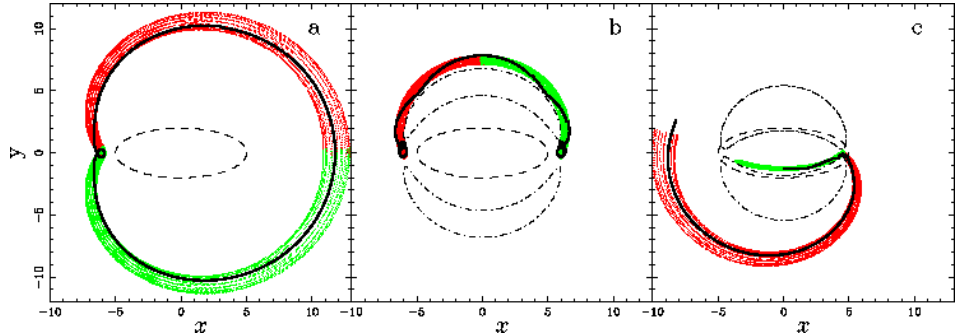}
\caption{Homoclinic {\bf (a)}, heteroclinic {\bf (b)} and escaping
{\bf (c)} orbits (black thick lines) in the configuration space. In
red lines, we plot the unstable invariant manifolds associated to the
periodic orbits, while in green we plot the corresponding stable invariant
manifolds. In dashed lines, we give the outline of the bar and, in 
{\bf (b)} and {\bf (c)}, we plot the zero velocity curves in dot-dashed lines.}
\label{fig:schtrans}
\end{figure*}

\begin{figure*}
\centering
\includegraphics[scale=0.5,angle=-90.]{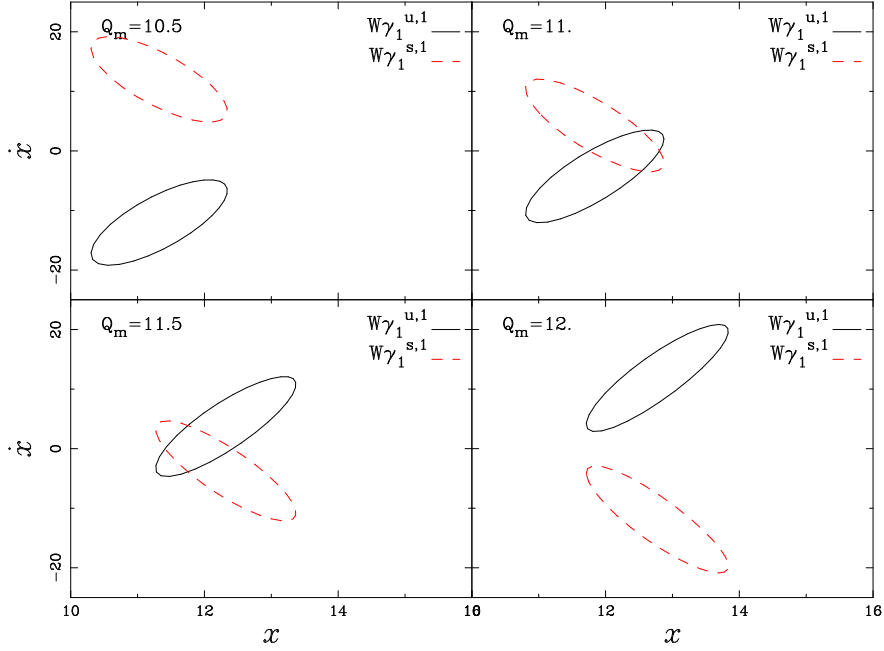}
\caption{Evolution of the curves $W_{\gamma_2}^{u,1}$ (black solid lines) and
$W_{\gamma_2}^{s,1}$ (red dashed line) on the phase space $x\dot x$ for a 
particular galactic model and four values of $Q_m$ (given in the upper left 
corner of each panel in units of $10^4$).}
\label{fig:seq}
\end{figure*}

\begin{figure*}
\begin{center}
\includegraphics[scale=0.5,angle=-90.]{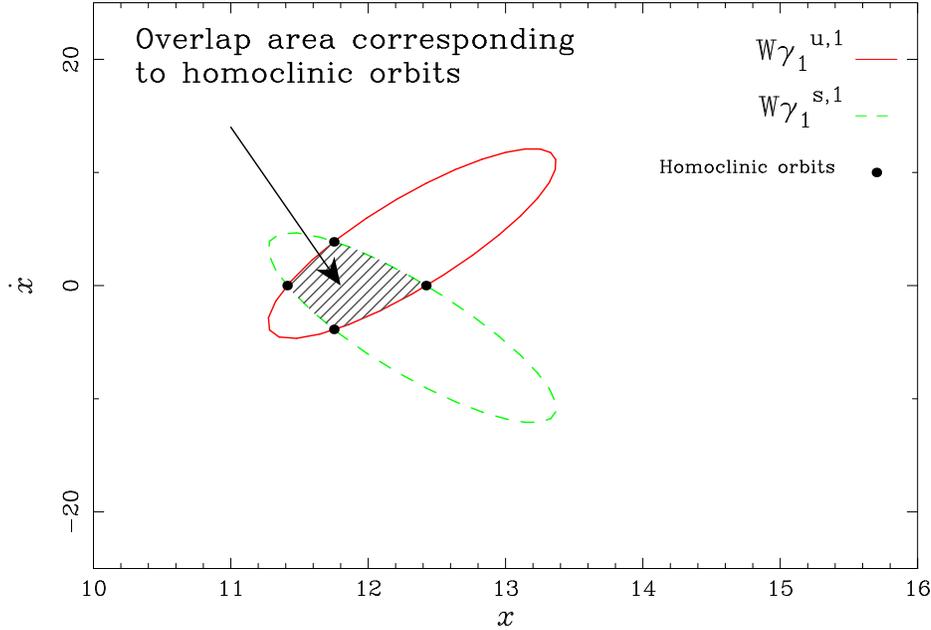}
\end{center}
\caption{Definition of the overlap area corresponding to homoclinic
orbits. The curves $W_{\gamma_2}^{u,1}$ (in red solid line) and
$W_{\gamma_2}^{s,1}$ (in green dashed line) intersect on the plane $(x,\dot
x)$. The four black points correspond to homoclinic orbits, while the
hatched area is the overlap area corresponding to homoclinic orbits.}
\label{fig:phs}
\end{figure*}

In our planar model, the dynamics take place in a four dimensional phase space. 
In fact, the number of dimensions can be reduced by one by fixing the energy to 
the energy level of the Lyapunov periodic orbit. In this three dimensional space, 
the stable and unstable invariant manifolds associated to a Lyapunov orbit are 
sets of asymptotic trajectories that form tubes. These tubes are filled and 
surrounded by a bundle of trajectories (see Fig~5 of Paper I, and G\'omez et al. 
\cite{gom04} and references therein for more details). A good way to visualise 
these tubes and make use of them is by means of Poincar\'e surfaces of section, this is, by 
drawing the crossings of the trajectories through a particular plane, or surface in 
phase space. Depending on the purposes of our study, some surfaces will be more 
suitable than others, but the methodology is the same. Let us take as an
example the surface of section $\cal{S}$ defined by $y=0$ with $x>0$, this is, 
we consider the orbits when they cut the plane $y=0$ having a positive value 
for the $x$ coordinate. Let us consider this surface of section $\cal{S}$ for the stable 
and unstable invariant manifolds of a Lyapunov orbit around $L_2$ (located in the $x<0$ side). 
Taking initial conditions close to the Lyapunov orbit and integrating $W^u_{\gamma_2}$ forward 
in time (resp. $W^s_{\gamma_2}$ backwards in time) until the first encounter with
$\cal{S}$ we obtain the simple closed curves $W^{u,1}_{\gamma_2}$ (resp. 
$W^{s,1}_{\gamma_2}$) which can be seen in Figs. \ref{fig:seq} and 
\ref{fig:phs} that will be commented later. In Fig.~\ref{fig:schtrans}a we can 
also see the plot of the trajectories until the encounter with 
$\cal{S}$. Although the simple closed curves $W^{u,1}_{\gamma_2}$ and 
$W^{s,1}_{\gamma_2}$
are obtained as the natural result of intersecting the manifold tubes with a 
plane, it should be mentioned that further crossings (i.e. $W^{u,k}_{\gamma_2}$
and $W^{s,k}_{\gamma_2}$ with $k>1$) may well not have this simple structure,
as can be seen in Gidea and Masdemont (\cite{gidmas}).

In the selected example, $W^{u,1}_{\gamma_2}$ and $W^{s,1}_{\gamma_2}$
are represented in $(x,\dot{x})$ coordinates. It is important to note that
a pair $(x,\dot{x})$ in $\cal{S}$ defines an orbit in a unique way, since $y=0$ 
and $\dot{y}$ is obtained from the energy level under study. By the definition of 
an invariant manifold, a point in
$W^{u,1}_{\gamma_2} \cap W^{s,1}_{\gamma_2}$ represents a trajectory
asymptotic to the Lyapunov orbit $\gamma$ around $L_2$ both forward and backward in
time. It is called an homoclinic orbit. In general, homoclinic orbits 
correspond to asymptotic trajectories, $\psi$, such that 
$\psi\in W^u_{\gamma_i}\cap W^s_{\gamma_i},\,i=1,2$. 
Thus, a homoclinic orbit departs asymptotically from 
the unstable Lyapunov periodic orbit $\gamma$ around $L_i$ and returns 
asymptotically to it (see Fig.~\ref{fig:schtrans}a). Heteroclinic orbits, on 
the other hand, are defined as asymptotic trajectories, $\psi^\prime$, such that
$\psi^\prime\in W^u_{\gamma_i}\cap W^s_{\gamma_j},\, i\ne j,\,i,j=1,2$. Thus, 
a heteroclinic orbit departs asymptotically from the periodic orbit 
$\gamma$ around $L_i$ and approaches asymptotically the corresponding Lyapunov 
periodic orbit with the same energy around the Lagrangian point at the opposite
end of the bar $L_j$, $i\ne j$ (see Fig.~\ref{fig:schtrans}b; a 
suitable surface of section for this computation can be the plane $x=0$). 

In our computations we also consider the overlap area corresponding 
to {\sl homoclinic} orbits as the area resulting from the intersection of 
the interior regions of $W^{u,1}_{\gamma_i}$ and $W^{s,1}_{\gamma_i}, \, i=1,2$
(see Fig.~\ref{fig:phs}). Analogously, we define the overlap area corresponding
to {\sl heteroclinic} orbits as the area resulting from the intersection  
of the interior regions of $W^{u,1}_{\gamma_i}$ and 
$W^{s,1}_{\gamma_j}, \, i\ne j,\,i,j=1,2$. 
$W^{u,1}_{\gamma_i}$ and $W^{s,1}_{\gamma_i}$ for a given model and energy may 
or may not intersect and, as the parameters of the model change, they can 
approach each other or move away. For instance, Fig.~\ref{fig:seq} shows the 
curves $W_{\gamma_2}^{u,1}$ and $W_{\gamma_2}^{s,1}$ on the plane {\bf $\cal{S}$}
for our model F, introduced in Sect.~\ref{sec:ferRFD},  for four different 
values of the quadrupole moment $Q_m$. We note that models with nonzero 
intersection of $W_{\gamma_i}^{u,1}$ and $W_{\gamma_i}^{s,1}$ are not 
associated to isolated values of the given free parameter (in our example 
$Q_m$), but they define a range. So we will find trajectories
that depart asymptotically from the periodic orbit, follow $W^u_{\gamma_i}$ and
do not intersect the corresponding stable invariant manifold in phase space. These 
trajectories initially spiral out from the region of the unstable periodic 
orbit (see Fig.~\ref{fig:schtrans}c) and we refer to them as escaping 
trajectories.

We will argue in Paper III that these three types of orbits -- namely, the 
homoclinic, the heteroclinic, and the escaping orbits -- become the backbone 
of ringed structures and of spiral arms observed in disc galaxies and we will 
follow how the overlap area gives a measure of these behaviours.

\section{Results}
\label{sec:res}
The next subsections are devoted to the results obtained when we vary the parameters 
of the models introduced in Sect. \ref{sec:model}. In order to best see the influence of 
each parameter separately, we make families of models in which only one of the free 
parameters is varied, while the others are kept fixed. We start by studying the effects 
of the main four parameters of model A. We then
study the effect of having reference models with a Ferrers bar potential and rising, flat, 
or falling rotation curves in the outer parts. Finally, we study the effects of having bar 
potentials with a larger influence in the outer parts. 

\subsection{Model A. The effect of the main four parameters}
\label{sec:fer01}
In this section we study the effect of the variation of the parameters of model A on the 
shape of the invariant manifolds. We consider both homogeneous ($n=0$) and inhomogeneous 
($n=1$) Ferrers bar models (Eq. \ref{eq:Ferden}). The main four parameters are the axial 
ratio of the bar, $a/b$, the Lagrangian radius, $r_L$, the quadrupole moment, 
$Q_m$, and the central concentration, $\rho_c$. For all four, we will take
the same range of values as in Athanassoula (\cite{ath92a}), since it is wide
enough to make sure we cover all relevant values. According to Kormendy (\cite{kor82}), 
typical bar axial ratios range from 2.5:1 to 5:1, while ovals have considerably smaller 
axial ratios. We thus choose values of the axial ratio within the range $a/b=1.2-6$. As the 
rings and spirals generally emanate from the ends of bars, the unstable Lagrangian points 
should be placed in the vicinity of the bar end points, which are at $x=\pm5$. We let the 
parameter $r_L$ vary accordingly within the range $r_L=3.5-8.5$, i.e. we consider Lagrangian 
points placed from well within the bar up to well beyond the bar end points. The values 
for the quadrupole moment are chosen within the range $Q_m=0.1\times 10^4-12\times 10^4$, covering 
all the range from weak bars or ovals to very strong bars. Finally, the central concentration 
is related to the presence of the spheroid. In this set of models the central
concentration is varied within the range $\rho_c=0.02\times 10^4-3.6\times 10^4$. For a given 
value of the bar central density, higher values of $\rho_c$ indicate denser bulges. 
We used two values of the concentration index of the bar, namely $n=0$ and $n=1$. For each of 
these indexes, we make families of models in which just one of the free parameters is varied, 
while the others are kept fixed. We use these families to study the influence of each of the 
free parameters on the global shape of the invariant manifolds.

\begin{figure}
$$
\includegraphics[scale=0.3,angle=-90.0]{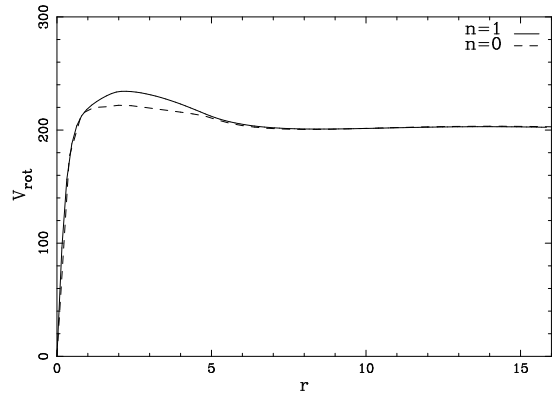}
$$
\caption{Rotation curves corresponding to model A for $n=0$ (dashed line) and 
$n=1$ (solid line).}
\label{fig:rot01}
\end{figure}

\begin{figure*}
\centering
\includegraphics[scale=0.97,angle=-90.0]{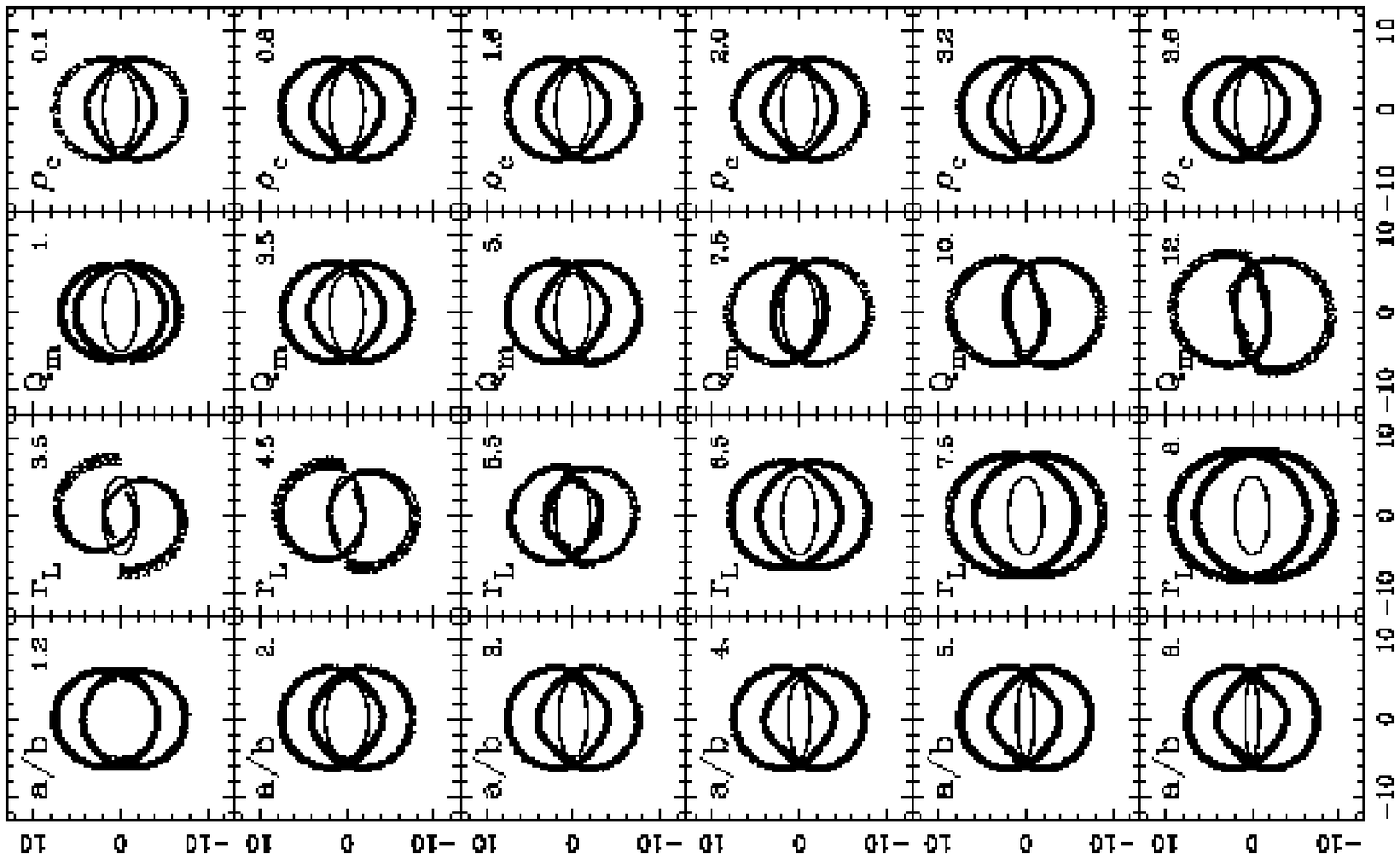} 
\caption{Ring and spiral structures in four sequences of models with
$n=0$ and various values of the four main parameters (given in the upper corners of
each panel). For the remaining parameters, see text. {\sl First column:} effect of 
the variation of the axial ratio of the bar, $a/b$; {\sl Second column:} effect of 
the variation of the Lagrangian radius, $r_L$; {\sl Third column:} effect of the 
variation of the quadrupole moment, $Q_m$ (in units of $10^4$); {\sl Fourth column:} 
effect of the variation of the central density, $\rho_c$ (in units of $10^4$). In all 
panels, we plot on the $x-y$ plane the outline of the bar (thin solid line) and the unstable
invariant manifolds associated to the Lyapunov orbits for a given
energy level.}
\label{fig:struct0}
\end{figure*}

In Fig. \ref{fig:struct0} we show the effect of the four free
parameters on the shape of the invariant manifolds for model A with $n=0$. In each
panel, we plot the outline of the bar and, for a given energy level, the unstable invariant
manifolds associated to the Lyapunov orbits of both $L_1$ and $L_2$ equilibrium points,
integrated until they perform half a revolution. For initial conditions of the invariant
manifolds very near the Lyapunov orbit, this time corresponds in all cases to 
approximately $3$ bar rotations, i.e. to $ \sim 0.25$ Gyr. In columns 1 to 4 we show the effect 
of the bar axial ratio (first column), of the Lagrangian radius (second column), of 
the quadrupole moment (third column) and of the central concentration (fourth
column). In all columns the numerical values of the parameters
increase from top to bottom. Results for $n=1$ are very similar, so
we do not show them here. 

Fig. \ref{fig:struct0} shows that the central concentration has hardly any influence on 
the shape of the invariant manifolds, which corresponds in all cases to the morphology of 
an $rR_1$ ringed galaxy. From the ends of the bar emanate branches that curl 
around the bar until the opposite end, forming two rings. One ring is close to
the bar and is elongated along it, but not far from circular, i.e. has the location and 
properties of observed inner rings. The other ring is at larger radii and is
elongated perpendicularly to the bar major axis, i.e. it has the location and properties
of an outer ring $R_1$. This description also holds for all values of the bar axial ratio.  
Varying this quantity has hardly any influence on the outline of the outer 
branches of the invariant manifolds, but does have some, albeit small, on the inner ones. 
Namely, for $a/b$ = 1.2, the inner ring is near-circular, while for larger 
values it is diamond shaped. 

On the contrary, varying the value of $r_L$ introduces different morphologies.
When the Lagrangian points are well within the bar, the inner ring is also 
within the bar, which is not realistic. The outer ring is not
closed, i.e. we have a spiral. However, this spiral also is not realistic,
because it does not emanate from the ends of the bar. For values of the Lagrangian 
radius beyond $r_L = 5$, the outer 
branches of the invariant manifolds tend to the opposite ends of the bar, and when 
they close they form an outer ring. Thus, the morphology is again that of an $rR_1$ 
ringed galaxy. As $r_L$ increases, the inner ring increases in size and becomes more 
circular. Already for $r_L = 6.5$ the inner ring size is considerably larger than the bar, 
contrary to what is observed. For the highest of the considered values, the inner ring forms a 
nearly circular structure with about double the size of the bar. 

Finally, the variation of the bar strength also gives different shapes of the ringed 
structures. In the case of weak bars, i.e. small $Q_m$ values, the inner ring is nearly 
circular with a diameter about the size of the bar. However, as will be discussed in 
paper III, the outline is not that of an $rR_1$ ringed galaxy. As the quadrupole moment 
increases and the bar gets stronger, the inner ring becomes more elongated along 
the bar until, for the case of strong bars, it is deformed in shape. The outer ring gets 
bigger in size with increasing $Q_m$, so that the ratio between the ring diameters increases. 
In the case of strong bars, both the inner and outer rings become asymmetric with respect 
to the major and minor axes of the bar and their major and minor diameters do not coincide 
any more with the $x$ and $y$ axes. 

\begin{figure*}[!ht]
\centering
\includegraphics[scale=0.55,angle=-90.0]{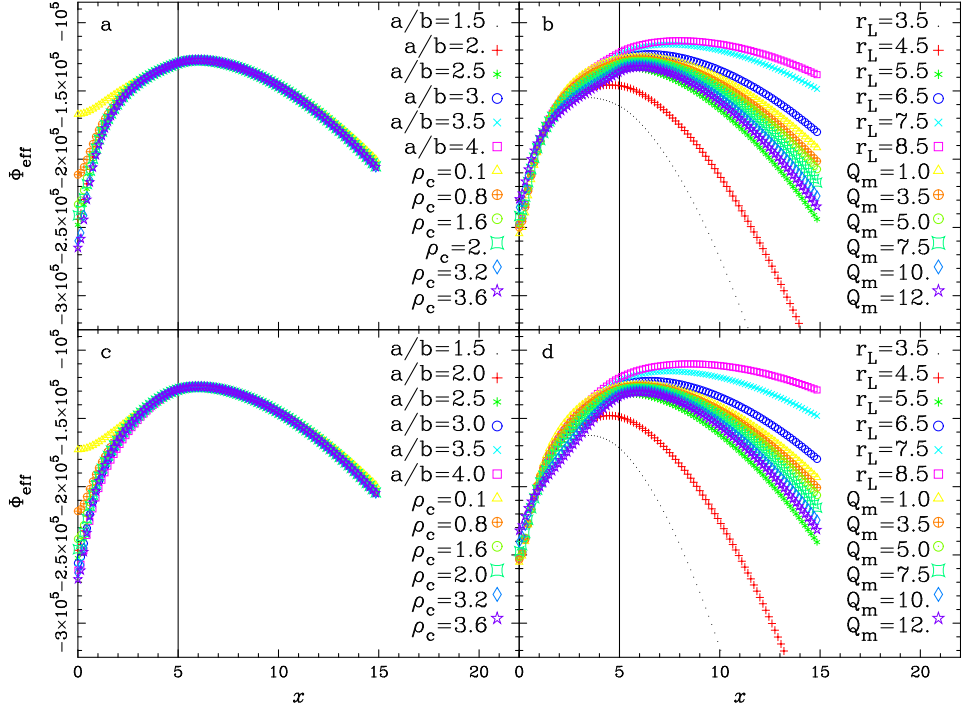}
\caption{Effective potential along the bar semi-major axis 
for model A with $n=0$ and $n=1$. {\bf (a)} Models with
homogeneous ($n=0$) bars with different values of
the bar axial ratio, $a/b$, and of the central density $\rho_c$ (in units of $10^4$). 
{\bf (b)} Models with homogeneous bars ($n=0$) with different values of
the Lagrangian radius, $r_L$, and of the quadrupole moment, $Q_m$ (in units of $10^4$). 
{\bf (c)} Same as in {\bf(a)} but with inhomogeneous bars ($n=1$). 
{\bf (d)} Same as in {\bf (b)} but with inhomogeneous bars ($n=1$). The
solid vertical line marks the end of the bar.}
\label{fig:effpot}
\end{figure*}

We can therefore conclude that the shape of the invariant manifolds is not sensitive to 
the variation of the axial ratio and of the central density. On the other hand, when we 
vary the Lagrangian radius or the quadrupole 
moment, we obtain richer structures. These differences are related to the values of 
the effective potential of the models. This can be seen in Fig. \ref{fig:effpot}, where 
we plot the effective potential of the models along the bar major
axis. First, we note that for radii smaller than the length of the bar, 
the effective potential shows some differences in all the cases. This means that 
all four free parameters will have an effect in the interior structure of the
bar and is in agreement with the fact that both the orbital structure and the gas flow
in the bar region vary considerably as the values of the four main parameters are changed
(Athanassoula \cite{ath92a}, \cite{ath92b}). Here, however, we are interested in the
spirals and rings which occur in a region from the ends of the bar outwards. For
radii larger than the length of the bar, the effective potential depends little
on the value of the axial ratio, or of the central concentration. This is illustrated 
in Fig. \ref{fig:effpot}a and Fig. \ref{fig:effpot}c for homogeneous and inhomogeneous 
bars, respectively. On the other hand, when we vary the Lagrangian radius or the 
quadrupole moment, the effective potential is considerably influenced. This is shown in 
Fig. \ref{fig:effpot}b and Fig. \ref{fig:effpot}d, again for homogeneous and inhomogeneous 
bars, respectively. This explains why only the variation of the Lagrangian radius and of the 
quadrupole moment has an effect on the global shape of the invariant manifolds.

\subsection{Models R, F and D. The effect of having a rising, flat, or falling rotation
curve with a Ferrers ellipsoid}
\label{sec:ferRFD}
In general, rotation curves in the outer parts of disc galaxies are
flat (Bosma \cite{bos81}). However, many cases of slightly rising or slightly
falling rotation curves are known. We thus want to study the
possible influences of these different rotation curves shapes on
the morphological structure of our models. 

\begin{figure}
\centering
\includegraphics[scale=0.3,angle=-90.0]{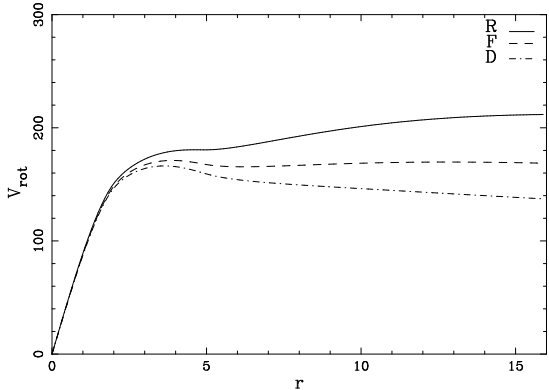}
\caption{Rotation curves corresponding to models R (solid line), F
(dashed line) and D (dot-dashed line). Note that all models have the
same rotation curve slope in the inner parts.}
\label{fig:rotABC}
\end{figure}

In this section we use reference models R, F and D, each having a different slope of 
the rotation curve in the outer parts. The rotation curves of these models are shown 
in Fig. \ref{fig:rotABC}. Note that in all cases the slope of the inner parts is the same, 
while in the outer parts it is rising (model R), flat (model F), or decreasing (model D).  
For each of these reference models we generate families of models, where we vary only one 
of the free parameters. Since in Sect. \ref{sec:fer01} we showed that the whole range of
two of the main parameters corresponds to the same morphology and dynamics beyond the bar 
region, and explained the reason for this, here we will study only the effect of the variation 
of the two crucial parameters, namely the values of $r_L$ and $Q_m$. The range of variation of 
these parameters will be the same as in Sect. \ref{sec:fer01}, i.e. $r_L=3.5-8.5$ and 
$Q_m=0.5 \times 10^4-12\times 10^4$, while the values of $\rho_c$ and $a/b$ are kept fixed to
$0.05\times 10^4$ and $2.5$, respectively.

\begin{figure*}
\centering
\includegraphics[angle=-90.0]{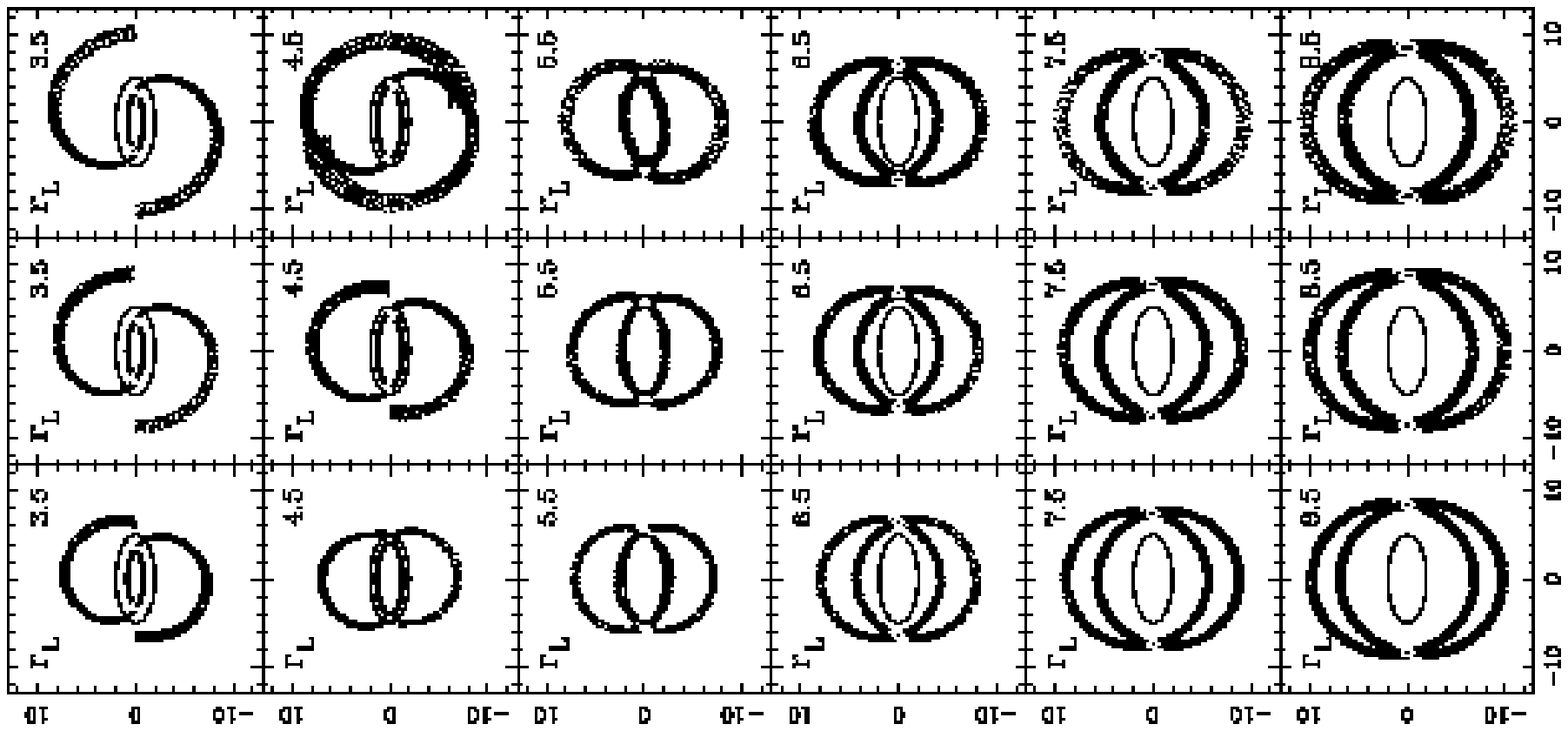}
\caption{Rings and spirals for three sequences of models with reference models R
(left), F (middle) and D (right) where we vary the
Lagrangian radius, $r_L$. This value increases from top to bottom and is given
in the upper right corner of each panel. The value of the quadrupole moment is
fixed to $Q_m=4.5\times 10^4$. In all panels, we plot on the $x-y$ plane the outline
of the bar (thin solid line), the two unstable Lyapunov orbits 
associated to the unstable equilibrium points $L_1$ and $L_2$ 
(white solid line), and the unstable invariant manifolds associated 
to the Lyapunov orbits, all for a given energy level.}
\label{fig:structABCrl}
\end{figure*}

In Fig. \ref{fig:structABCrl} we show the effect of the variation of
the Lagrangian radius on the shape of the invariant manifolds. 
In all columns we increase the value of $r_L$ from top to bottom, while the value
of the quadrupole moment is fixed to $Q_m=4.5 \times 10^4$. We note that,
for rotation curves which in the outer parts are rising or flat, 
the shape of the invariant manifolds presents the same behaviour as in model 
A (see previous subsection). Thus, for small $r_L$ values we have spiral arms, while for 
large $r_L$ values we have outer rings. An inner ring is present in all cases and its
 size increases with $r_L$. It is smaller than the bar, i.e. unrealistic, for $r_L < a$ and 
much larger than the bar, i.e. again unrealistic, for very large $r_L$ values, while realistic
sizes are only found for an intermediate range of values. In the case of falling rotation curves, 
the evolution is similar. However, when the Lagrangian radius is close to the bar length, 
the invariant manifolds cross each other, forming an $R_2$ ring. 

Along each row we observe the effect of changing the slope of the rotation curve at the 
outer radii, while keeping constant the value of all bar parameters, including the corotation 
radius. In the upper row, where the Lagrangian points are inside the bar, the spiral arms 
open as the slope of the rotation curve decreases. A case where corotation is near the ends 
of the bar is shown in the second row. A rising rotation curve gives an $rR_1$ ring, while
a flat rotation curve gives a $R_1$ pseudo-ring. In the specific
case of a falling rotation curve, the outer branches of the invariant
manifolds are long and open, finally crossing each other, thus forming an $R_2$ ring. 
For values of the corotation radius beyond the end of the bar, the shape 
corresponds to that of an $rR_1$ ring galaxy.
 
\begin{figure*}
\centering
\includegraphics[angle=-90.0]{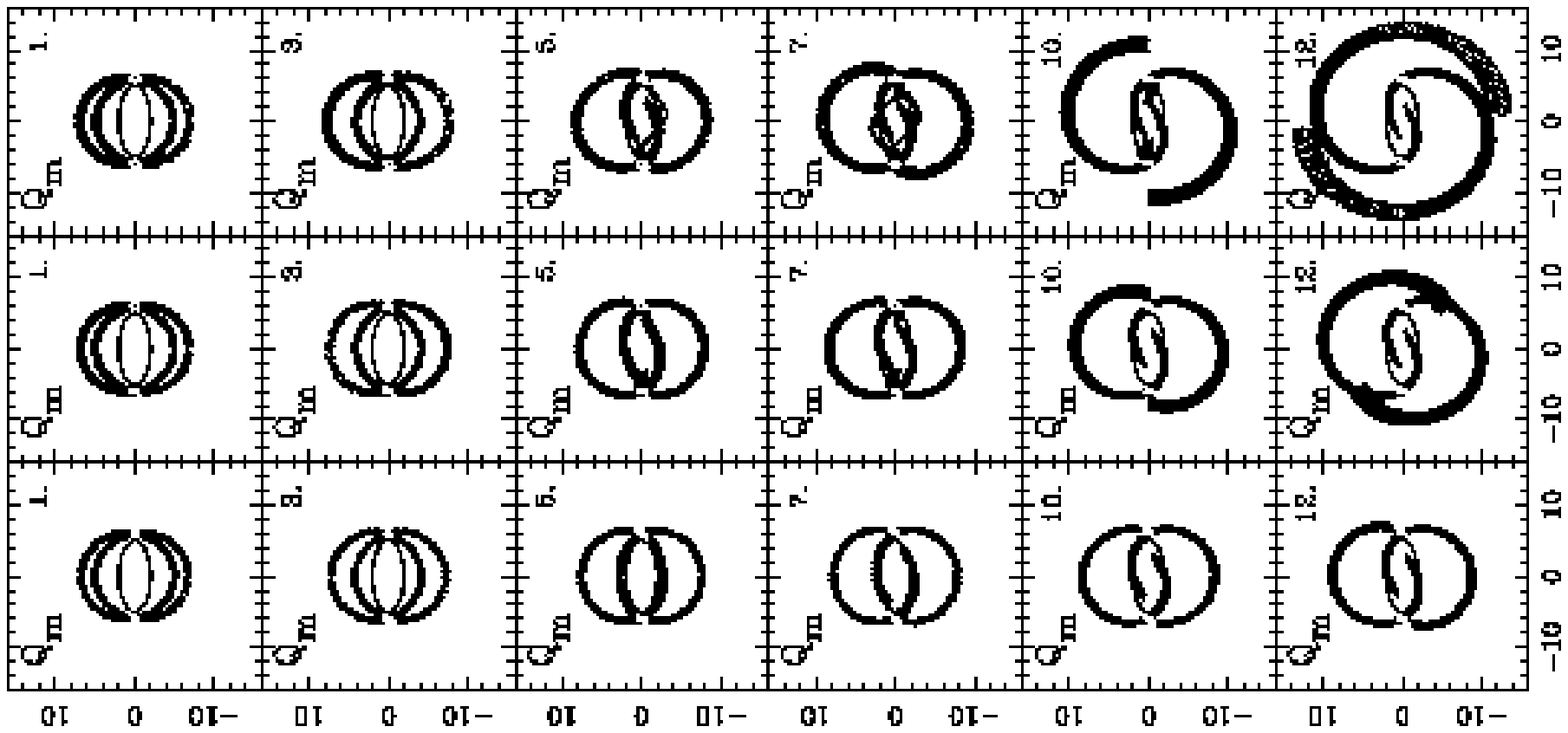}
\caption{Same as in Fig. \ref{fig:structABCrl}, for three sequences of models
with reference models R (left), F (middle) and D (right) and different values
of the quadrupole moment, $Q_m$, given in units of $10^4$ in the upper right corner
of each panel. The value of the Lagrangian radius is fixed to $r_L=6$.}
\label{fig:structABCqm}
\end{figure*}

In Fig. \ref{fig:structABCqm}, we show the effect of the variation of the quadrupole moment 
on the shape of the invariant manifolds. In all columns we increase the value of $Q_m$ from 
top to bottom, while the value of the Lagrangian radius is fixed to $r_L=6$. In the case of 
reference models with rising or flat rotation curves (first and second columns), the behaviour 
of the invariant manifolds is the same as in model A. For a falling
rotation curve and a weak bar, we still have an $rR_1$ ring structure. As we increase 
the bar strength, first the inner ring becomes distorted, then the major axis of the 
outer ring ceases to be perpendicular to the bar major axis, and finally, with strong 
bars, the outer invariant manifolds either do not close and form spiral arms, or they 
are long and cross each other, thus forming $R_2$ rings. As in the previous figure, along 
each row we observe the effect of changing the slope of the rotation curve in 
the outer parts, while keeping constant the bar quadrupole moment. For weak and 
intermediate strength bars, the shape of the rotation curve does not influence the 
morphology. For strong bars, however, we obtain different structures. A rising
rotation curve gives an $rR_1$ ring, while a flat rotation curve gives an 
asymmetric ring. In the case of falling rotation curves, the outer branches of 
the invariant manifolds are open and form spiral arms and $R_2$ rings.

\subsection{Models R', F' and D'. Other bar models}
\label{sec:dehRFD}
\begin{figure}
$$
\includegraphics[scale=0.3,angle=-90.0]{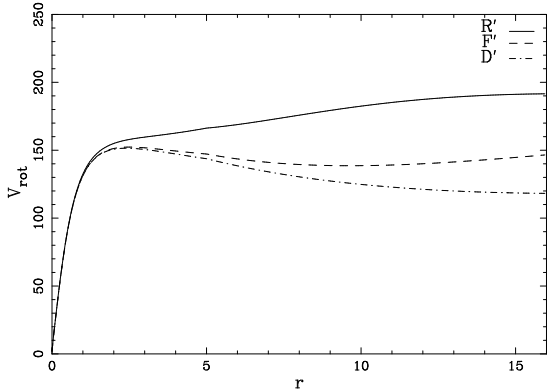}
$$
\caption{Rotation curves corresponding to models R' (solid line), F'
(dashed line) and D' (dot-dashed line). Note that all models have the
same rotation curve slope in the inner parts.}
\label{fig:velrotDEF}
\end{figure}

As previously mentioned, Ferrers ellipsoids do not have a great influence
in the outer region, which is where the outer branches of the invariant manifolds 
are found. Since we argue in this paper that invariant manifolds are responsible for 
the rings and spiral structures, we considered also bar potentials of either of Dehnen 
type (Eq. \ref{eq:adhoc1}), or of Barbanis-Woltjer type (Eq. \ref{eq:adhoc2}). 
The results of the two models are, however, very similar, so for the sake of brevity we 
discuss only the results for the Dehnen type bar here.
In this case we use models R', F' and D', again each with a different slope of the 
rotation curve in the outer parts. The rotation curves of these models are shown in
Fig. \ref{fig:velrotDEF}. As in the previous sections, we make families of models, varying 
one of the parameters, while keeping the rest constant. Since these models are ad hoc and 
they do not correspond to any simple and realistic density, choosing the parameters is 
less easy, since there are less constraints from observations. Based on the results of 
Sect. \ref{sec:fer01}, we use as free parameters only the Lagrangian radius, $r_L$, and 
$\epsilon$, which determines the bar strength. The values of $r_L$ are taken within the 
range $r_L=3.5-8.5$, as in the previous sections. The values of $\epsilon$ are taken within 
the range $\epsilon=0.01-0.3$, which corresponds a range of $5$ to $26\%$ for the ratio of 
the non-axisymmetric component to the axisymmetric component of the force in the $x$-direction 
and at the radius $r=16$.

\begin{figure*}
\centering
\includegraphics[angle=-90.0]{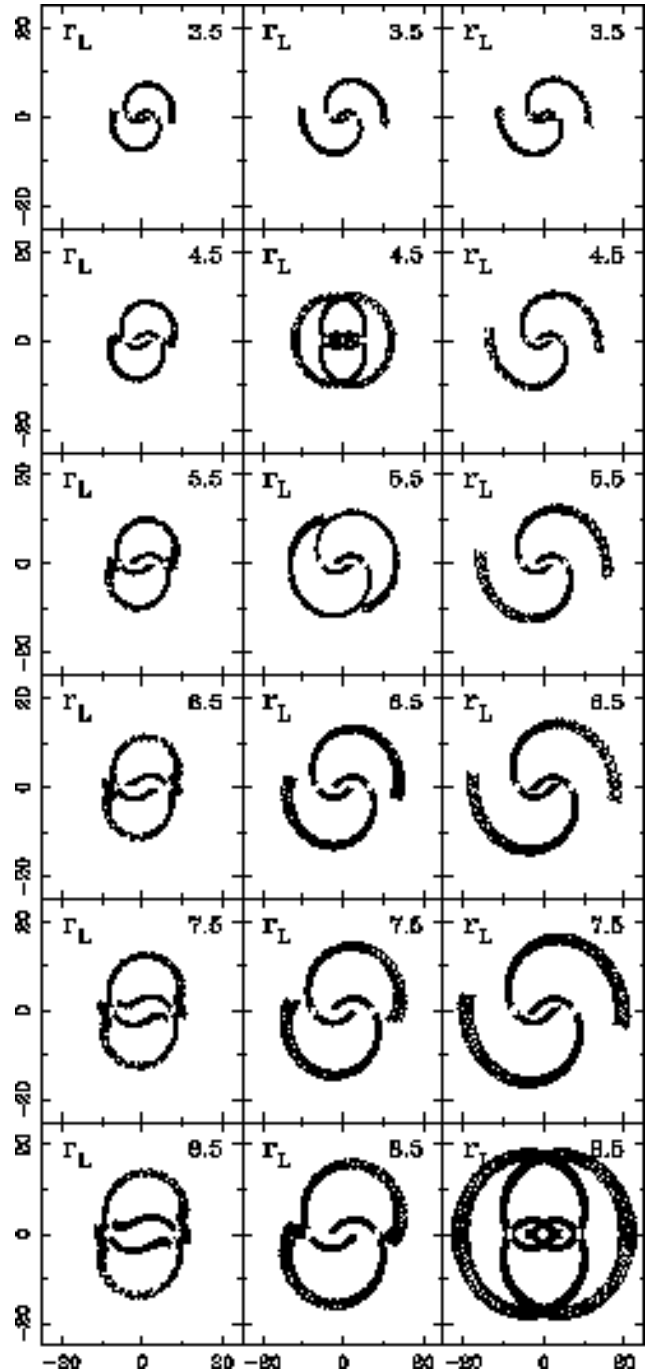}
\caption{Rings and spirals for three sequences of models with reference models R' (left),
F' (middle) and D' (right), where we vary the Lagrangian radius, $r_L$. 
This value increases from top to bottom and is given in the upper right corner of each
panel. The value of the strength parameter is fixed to $\epsilon=0.15$. In all panels, we 
plot on the $x-y$ plane the two unstable Lyapunov orbits associated to the
unstable equilibrium points $L_1$ and $L_2$, and the unstable invariant manifolds associated 
to the Lyapunov orbits, all for a given energy level. For model F' and $r_L=4.5$ and model D' 
and $r_L=8.5$, we plot the corresponding stable invariant manifolds.}
\label{fig:structDEFr}
\end{figure*}

In Fig. \ref{fig:structDEFr} we show the effect of the Lagrangian radius 
on the shape of the invariant manifolds. In all columns we increase the value of $r_L$ from 
top to bottom, while the strength parameter $\epsilon$ is fixed to $0.15$. For model R', i.e. 
for a rotation curve which is slightly increasing, we start at the smallest $r_L$ with a spiral. 
As the value of $r_L$ is increased the spiral gets more tightly wound and then closes forming a 
somewhat asymmetric outer ring. These changes are accompanied by an increase of the size of 
the inner ring, which, nevertheless, is in all cases quite asymmetric. For model D', i.e. for 
a somewhat decreasing rotation curve, at the smallest $r_L$ value we have a rather open spiral. 
As the value of $r_L$ is increased, the extent of the spiral increases considerably and its pitch 
angle increases slightly. For the largest $r_L$ value, $r_L=8.5$, the outer branches of the 
stable and unstable invariant manifolds delineate well the loci of both types of outer rings, 
namely $R_1$ and $R_2$, thus forming an $R_1R_2$ ring. The inner ring also increases in size 
as the value of $r_L$ increases, while staying of roughly the same shape. The position angle 
leads that of the bar by an angle which, as we increase $r_L$, increases approximately from 
$20$ to $36$ degrees. The sequence with the flat rotation curve, F', is the richest in 
morphological types. It starts with a rather open spiral, as the D' sequence, but, contrary 
to the D', the shape of the invariant manifold branches changes considerably. Thus, for the 
largest $r_L$ value they form asymmetric outer and inner rings, similar in outline to those 
formed for model R' and $r_L=4.5$. In between these two, the invariant manifolds give us 
two different morphologies. For $r_L=4.5$, the outer branches of the stable and unstable 
invariant manifolds outline the shape of $R_1R_2$ rings, as for model D' and $r_L=8.5$, 
while the inner ring is deformed in shape. For $r_L=5.5$, the outer branches of the unstable 
invariant manifolds evolve until they cross each other forming an outer ring whose principal 
axis is parallel to the bar major axis, thus forming an $R_2$ outer ring. 

In each row we observe the effect of keeping constant the value of corotation radius and of
all other bar parameters, while changing the slope of the rotation curve at outer radii. 
When the Lagrangian radius is smaller than the bar length, the spiral arms open as the slope 
of the rotation curve decreases. The case where corotation is approximately the value of the 
bar length scale is shown in the second row. A rising rotation curve gives a broken and 
misaligned $rR_1$ ring, while a flat rotation curve gives $R_1R_2$ rings, and a falling 
rotation curve gives spiral arms. The tendency is similar if the values of the Lagrangian 
radius are somewhat larger than the bar length scale. For a rising rotation curve, the 
structure of the galaxy is that of an $rR_1$ ring. However, if the rotation curve is flat, 
we obtain $R_2$ rings, while if the rotation curve is falling, we obtain spiral arms. As we 
increase the value of the Lagrangian radius further, the invariant manifolds tend to approach 
the opposite ends finally forming outer rings misaligned with respect to the bar. The inner 
rings, in turn, tend to open as the Lagrangian radius increases. In the case of a falling 
rotation curve, the invariant manifolds form spiral arms. 
 
\begin{figure*}
\centering
\includegraphics[angle=-90.0]{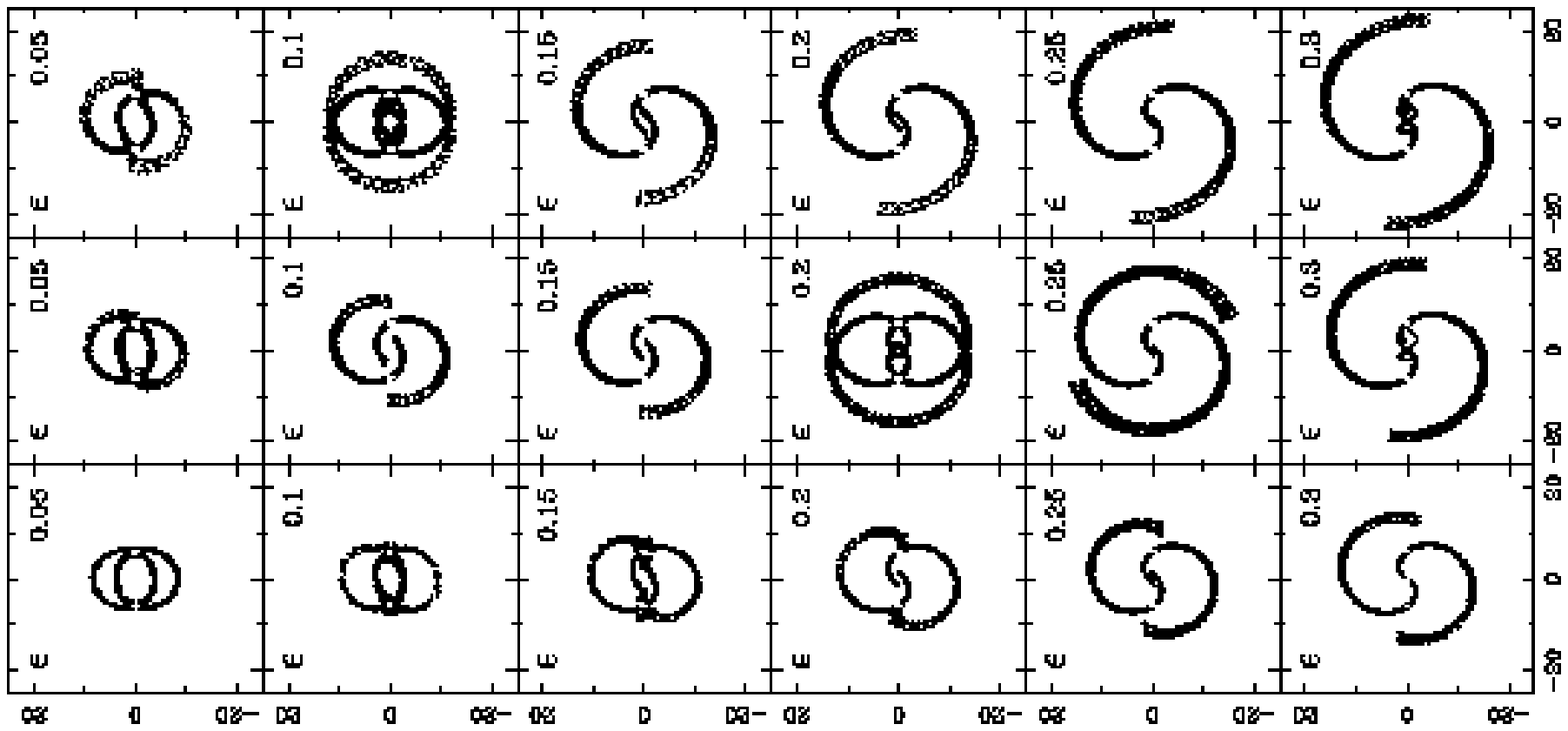}
\caption{Same as in Fig. \ref{fig:structDEFr} for three sequences of models
with reference models R' (left), F' (middle) and D' (right) and different values of the 
bar strength parameter, $\epsilon$. The value of the Lagrangian radius is fixed to $r_L=6$.
For model F' and $\epsilon=0.2$ and model D' and $\epsilon=0.1$, we plot the corresponding
stable invariant manifolds.}
\label{fig:structDEFf}
\end{figure*}

In Fig. \ref{fig:structDEFf}, we show the effect of the bar strength on the shape of the 
invariant manifolds. In all columns we increase the value of $\epsilon$ 
from top to bottom, while the value of the Lagrangian radius is fixed to $r_L=6$.
In the case of reference models with a rising rotation curve, the invariant manifolds tend 
to open. Thus, we have a behaviour opposite to the one we found when we increased the value 
of the Lagrangian radius. That is, for low values of the parameter $\epsilon$, the
galaxy morphology is that of an $rR_1$ ring, while as the bar strength increases, the branches
of the invariant manifolds tend to open and form spiral arms. The inner ring is,
for low values of $\epsilon$, symmetric. As the bar strength increases, it becomes more 
eccentric and finally asymmetric. For a flat rotation curve and a weak bar, we 
initially have an $rR_1$ ring structure. As we increase the bar strength, the inner
ring becomes distorted, while the major axis of the outer ring ceases to be perpendicular 
to the bar major axis. Increasing the bar strength, we find spirals. A further increase of the 
bar strength brings a morphology of the $R_1R_2$ type (see middle panel for $\epsilon=0.2$) and
then (see middle panel for $\epsilon=0.25$) a morphology of the $R_2$ type. Finally, with strong 
bars the outer invariant manifolds do not close, thus forming spiral arms. For a falling rotation 
curve, the behaviour is similar to the F' case but the invariant manifolds become open for 
weaker bars. As the bar strength increases, the outer branches of the invariant manifolds 
open forming $R_1R_2$ rings and spiral arms, but the inner branches will appear distorted. 

In each row, we observe the effect of changing the slope of the rotation curve in the outer 
parts. For weak bars, the invariant manifolds tend to open as the slope of the rotation 
curve decreases. Note that this behaviour is similar to the one obtained for keeping constant 
the value of corotation radius. We obtain $R_1R_2$ rings for weak bars and falling rotation 
curves or intermediate strength bars and flat rotation curves. In the case of strong bars, 
the morphologies obtained are spiral arms and $R_2$ rings.

\section{Summary and discussion}
\label{sec:con}
In this paper we developed an idea originally proposed in Paper I, namely that spirals and 
rings in barred galaxies can be explained in a common dynamical framework. This is based on 
the dynamics of the unstable equilibrium points in the rotating bar potential and, more
specifically, on the invariant manifolds associated to the unstable Lyapunov periodic orbits 
around these equilibrium points. These unstable Lyapunov orbits are simple orbits of 
elliptical-like shape, encircling one of the unstable equilibrium points and staying always in 
its immediate vicinity. The dynamics of the corresponding manifolds have been well studied by 
orbital structure theory and we have drawn heavily from that field for our purposes. It is not, 
however, necessary to master the intricacies of the definition and dynamics of invariant 
manifolds in order to follow our work. The reader may simply think of a set (a bundle) of 
orbits emanating from the vicinity of the Lyapunov orbits, i.e. from nearby positions and 
velocities.

We investigated different barred galaxy models, to make sure that none of our principal results 
are model dependent and also in order to give ourselves a broad base to allow us, in Paper III, 
to discuss globally the application of our results to real galaxies. Model A is identical to 
that introduced by Athanassoula (1992a). Models R, F and D are similar, but allow for rotation 
curves which can be slightly rising (R), flat (F), or decreasing (D). Finally, Models R', F' 
and D' (or R" and D") are similar to R, F and D, but have a Dehnen-type (or a Barbanis-Woltjer 
type) bar.

\begin{figure*}
\centering
\includegraphics[scale=0.19,angle=-90.0]{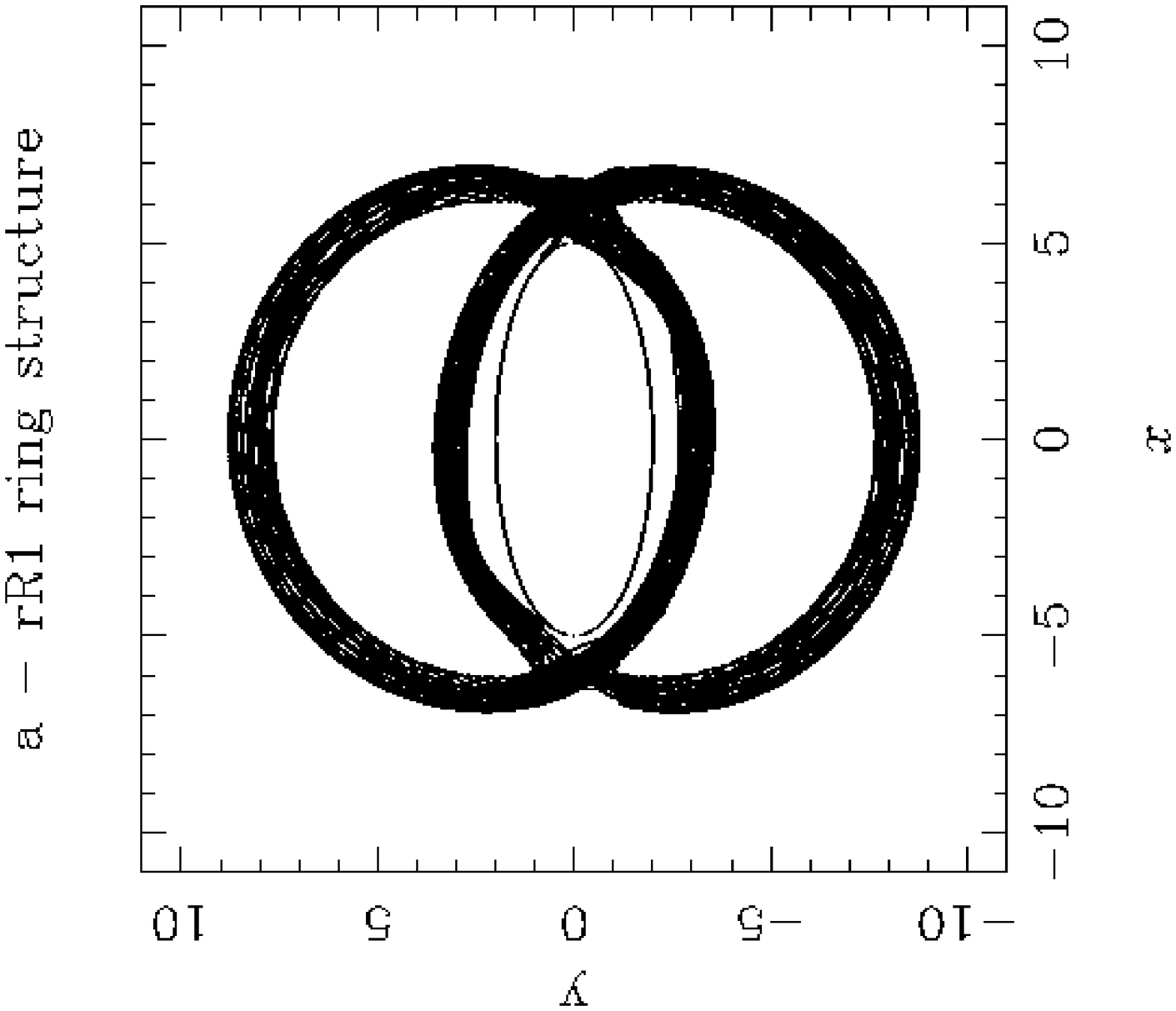}\hspace{0.15cm}
\includegraphics[scale=0.19,angle=-90.0]{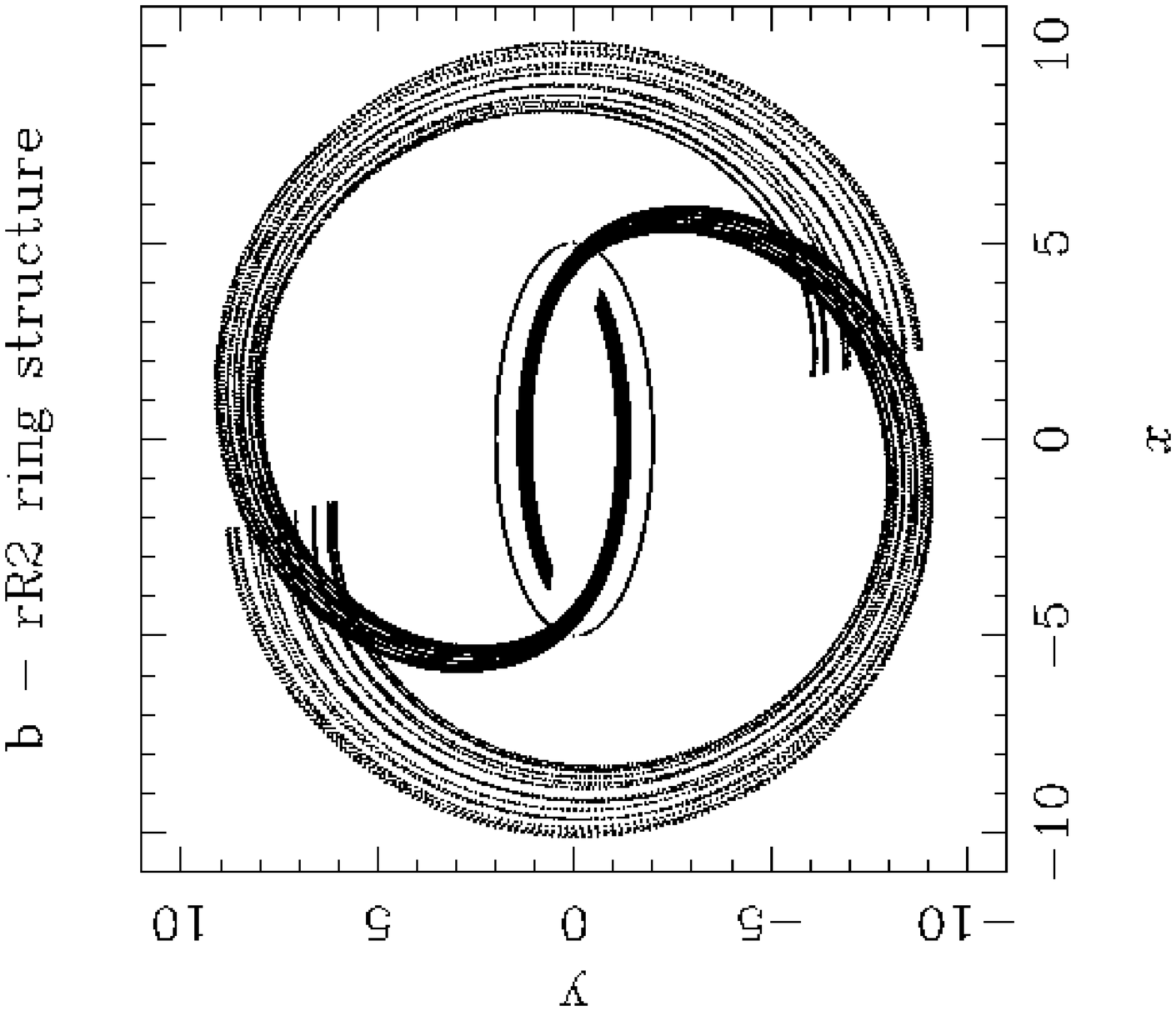}\hspace{0.15cm}
\includegraphics[scale=0.19,angle=-90.0]{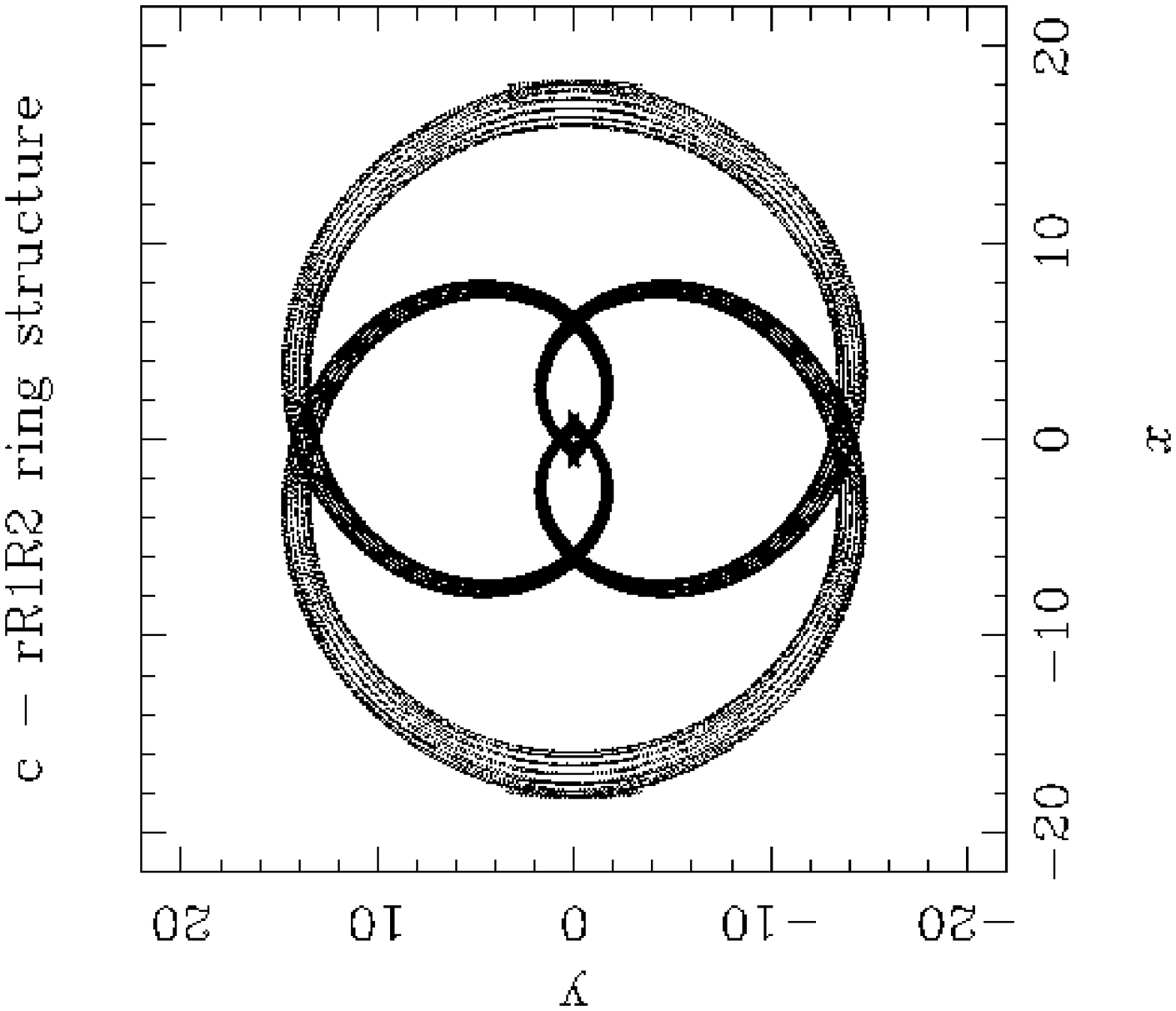}\hspace{0.15cm}
\includegraphics[scale=0.19,angle=-90.0]{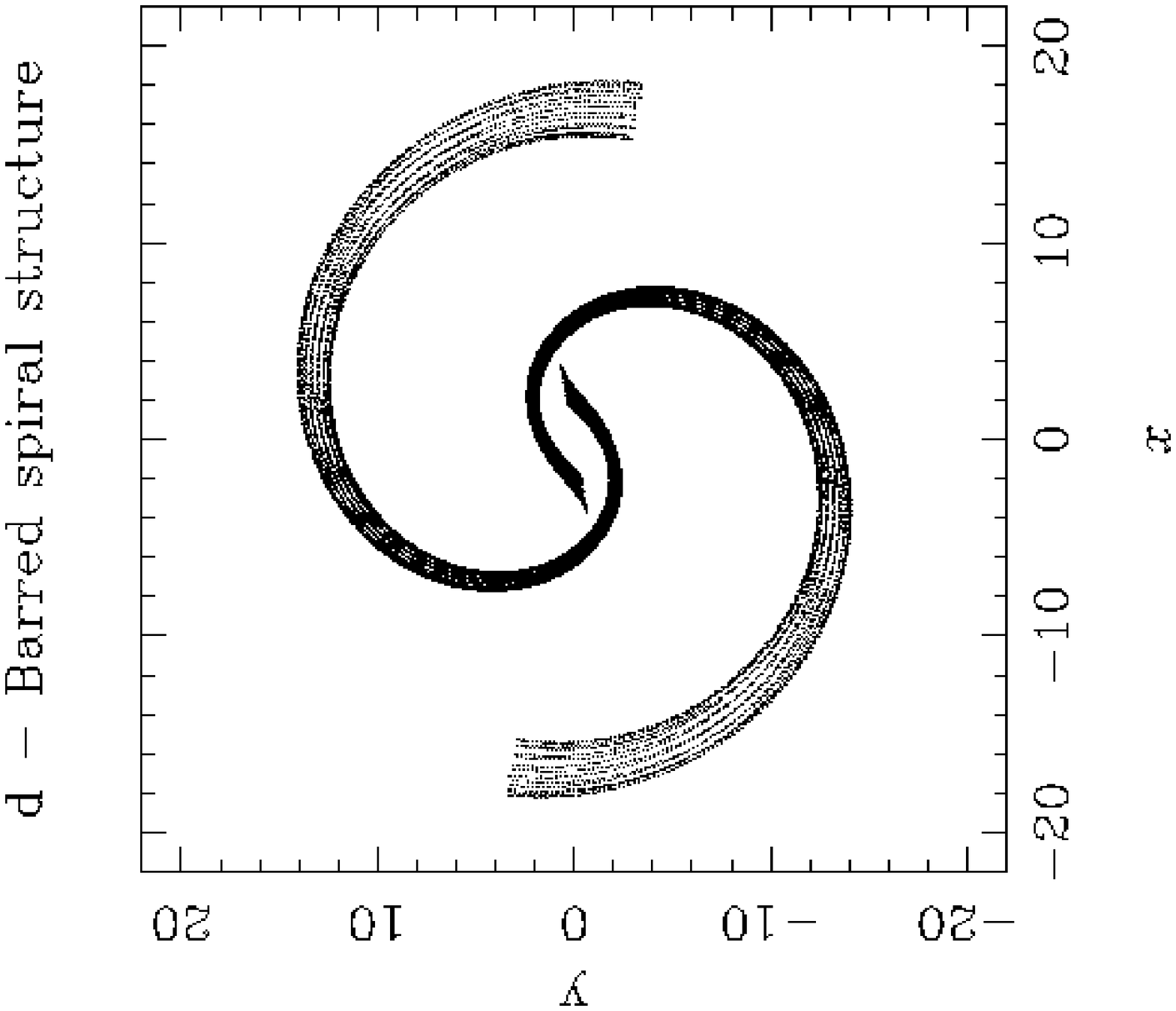}
\caption{Rings and spiral arms structures. We plot the invariant manifolds for different
models. {\bf (a)} $rR_1$ ring structure with parameters: Model D with a Ferrers
ellipsoid with $n=1$, $a/b=2.5$, $r_L=6$, $Q_m=4.5$, and $\rho_c=0.05$. {\bf (b)} 
$rR_2$ ring structure with model D, a Ferrers ellipsoid with $n=1$, $a/b=2.5$, $r_L=4.5$, 
$Q_m=4.5$, and $\rho_c=0.05$. {\bf (c)} $R_1R_2$ ring structure with model F' with a generalized
Dehnen's bar potential, $\epsilon=0.24$, $r_L=6$. {\bf (d)} Barred spiral galaxy with model D' 
with a generalized Dehnen's bar potential, $\epsilon=0.15$, $r_L=6$.}
\label{fig:family}
\end{figure*}

Our main result is that the loci outlined by the invariant manifolds and by the orbits 
associated with them can reproduce, for appropriate values of the model parameters, all the 
morphologies observed in real galaxies. Thus we can obtain two spiral arms emanating from 
the ends of the bar, or rings. We obtain inner rings as well as all the varieties of outer 
rings, namely $R_1$, $R_2$ and $R_1R_2$. This is shown in Fig.~\ref{fig:family}, where we 
show one typical example of each kind of morphology. These include an $rR_1$ morphology 
(Fig.~\ref{fig:family}a) as in NGC~2665 (Buta \& Crocker \cite{but91}), an $rR_2$ one 
(Fig.~\ref{fig:family}b) as in ESO~325~-~28 (Buta \& Crocker \cite{but91}), an $R_1R_2$ 
one (Fig.~\ref{fig:family}c) as in NGC~3081 (Buta \& Purcell \cite{but98}; Buta, Byrd \& 
Freeman \cite{but04}), and a spiral (Fig.~\ref{fig:family}d) as in NGC~1365 
(J\"{o}rs\"{a}ter \& Moorsel \cite{jor95}). It is important to stress that all these types 
were obtained with the {\it same} dynamics. Indeed, all that differs between these four 
examples is the values of the model parameters and the bar model potential.

Model A has four basic free parameters describing the bar dynamics. We show that two of 
these, the bar axial ratio and the central concentration of the model, hardly influence 
the loci of the invariant manifolds, while the two others, namely the Lagrangian radius 
and the bar quadrupole mass, have a considerable effect. We explain this by looking at the 
influence of these four parameters on the effective potential of the system. The main type of
morphology found in model A is the $rR_1$ type. Allowing for different slopes of the rotation 
curve and/or different bar models, we introduce the remaining morphological varieties.

In this paper we considered a very large number of models and of model parameters and 
calculated in all cases the appropriate invariant manifolds in order to find the resultant 
morphology. This can also be applied to real galaxies. Using an image of the galaxy in 
the near infrared and with an estimate of the thickness and of the mass-to-light ratio, 
one can obtain the bar potential. Including information from the rotation curve, one can 
obtain the total potential in which the manifolds and orbits can be calculated. A 
similar procedure has been already applied e.g. by Lindblad, Lindblad \& Athanassoula 
(\cite{lin96}) to study the gas flow in NGC~1365 and by Byrd, Buta \& Freeman (\cite{byr06}) 
to study the rings of NGC~3081. The latter work in particular studies the shape of the inner 
and outer rings both analytically and with the help of simulations. Of course, as is the case 
for all modelling of observations, e.g. concerning the mass-to-light ratios and the thickness, 
this procedure relies on a few approximations and assumptions. Yet it has 
revealed a lot of important information on bars and spirals (e.g. Sanders \& Tubbs \cite{san80};
Duval \& Athanassoula \cite{duv83}; Lindblad, Lindblad \& Athanassoula \cite{lin96}; 
Patsis, Athanassoula \& Quillen \cite{pat97}; Weiner, Sellwood \& Williams \cite{wei01};
Kranz, Slyz \& Rix \cite{kra03}; Perez, Fux \& Freeman \cite{per04}; Byrd, Buta \& Freeman
\cite{byr06}) and would thus be well worth pursuing for the model presented here.

This paper leaves two major points unanswered. The first one is why a given set of model 
parameters gives a given morphology? Our work so far shows that this is not a random process, 
but does not explain the physics behind it. Thus we are not able, at this point, to foresee what
morphology will result from each particular bar model and set of specific parameters without 
calculating the necessary invariant manifolds. The second unanswered point is to what extent 
the results found here are relevant to real galaxies. We have of course shown that the loci of 
the appropriate invariant manifolds can reproduce all the observed morphologies. This is very 
encouraging, but does not, on its own, assure us that we have found here the theory that explains
the formation of ring and spiral structures in barred galaxies. Several more points need to be 
considered before this conclusion can be reached. These yet unanswered two points will be
addressed in Paper III.

{\bf Acknowledgements}
We thank Albert Bosma for stimulating discussions of the properties of
observed rings. This work is supported by the Spanish MCyT-FEDER Grant 
MTM2006-00478. MRG acknowledges her ``Becario MAE-AECI''.

\end{document}